# Non-Bloch-Siegert-type power-induced shift of two-photon electron paramagnetic resonances of charge-carrier spin states in an OLED

S. I. Atwood[1], S. Hosseinzadeh[1], V. V. Mkhitaryan[2], T. H. Tennahewa[1], H. Malissa[1,2], W. Jiang[3], T. A. Darwish[4], P. L. Burn[3], J. M. Lupton[1,2], and C. Boehme[1]

[1]Department of Physics and Astronomy, University of Utah, Salt Lake City, Utah 84112, USA

[2]Institut für Experimentelle und Angewandte Physik, Universität Regensburg, Universitätsstrasse 31, 93053 Regensburg, Germany

[3]Centre for Organic Photonics & Electronics, School of Chemistry and Molecular Biosciences, The University of Queensland, Brisbane, QLD 4072, Australia.

[4]National Deuteration Facility, Australian Nuclear Science and Technology Organization (ANSTO), Lucas Heights, NSW 2234, Australia.

## Abstract

We present Floquet theory-based predictions and electrically detected magnetic resonance (EDMR) experiments scrutinizing the nature of two-photon magnetic resonance shifts of charge-carrier spin states in the perdeuterated π-conjugated polymer poly[2-methoxy-5-(2'-ethylhexyloxy)-1,4-phenylene vinylene] (d-MEH-PPV) under strong magnetic resonant drive conditions (radiation amplitude $B_1$ ~ Zeeman field $B_0$). Numerical calculations show that the two-photon resonance shift with power is nearly drive-helicity independent. This is in contrast to the one-photon Bloch-Siegert shift that only occurs under non-circularly polarized strong drive conditions. We therefore treated the Floquet Hamiltonian analytically under arbitrary amplitudes of the co- and counter-rotating components of the radiation field to gain insight into the nature of the helicity dependence of multi-photon resonance shifts. In addition, we tested Floquet-theory predictions experimentally by comparing one-photon and two-photon charge-carrier spin resonance shifts observed through room-temperature EDMR experiments on d-MEH-PPV-based bipolar injection devices [i.e, organic light emitting diode structures (OLEDs)]. We found that under the experimental conditions of strong, linearly polarized drive, our observations consistently agree with theory, irrespective of the magnitude of $B_1$, and therefore underscore the robustness of Floquet theory in predicting nonlinear magnetic resonance behaviors.

## 1. Introduction

Multi-photon transitions of magnetic dipoles and drive-induced resonance-peak shifts are hallmark phenomena of magnetic resonance in the strong-coupling limit, and are characterized by conditions where the magnetic resonant drive field amplitude, $B_1$, is comparable in magnitude to the static magnetic Zeeman field, $B_0$ [1-3]. Historically, Bloch and Siegert pioneered the description of the drive-induced resonance shift for one-photon resonances [3], highlighting this as a drive-field and helicity-dependent phenomenon. Subsequent work has delved into multi-photon conditions under a range of limiting approximations [4-7], with studies particularly focusing on scenarios where both two-photon transitions and magnetic resonance peak center shifts are evident under strong drive. For purely Zeeman-split electron spin states, Figure 1 showcases both (a) the two-photon transition and (b) a drive field-induced resonance shift. Early qualitative indications for shifts in two- and three-photon magnetic dipole transitions emerged from studies on optically pumped sodium by Margerie and Brossel [7]. Shortly after that observation, a theory explaining these transitions was put forward by Winter [8] and further developed by Cohen-Tannoudji and Haroche [9,10]. A qualitatively different observation technique was utilized by Morozov *et al.*, who reported findings for qualitative two- and three-photon shifts in radical-ion pairs of aromatic acceptors in liquids using optically detected magnetic resonance (ODMR) experiments [8]. Recently, Sun *et al.* conducted optical magnetic resonance measurements on $^{133}$Cs atoms and quantitatively analyzed two- and three-photon resonance shifts [12]. Similarly, Fregenal *et al.* observed peak shifts in many higher multi-photon transitions while studying $n$-photon resonances, with $n$ reaching up to 23 for high orbital angular momentum states in Li Rydberg atoms through detection by selective field ionization [13]. In recent years, the exploration of strong-drive induced electron spin resonance shifts using electrically detected magnetic resonance (EDMR) has gained traction. Analogous to ODMR, EDMR enables the observation of electron spin resonances in scenarios of minimal thermal spin polarization. This is achieved by detecting the spin-permutation symmetry of weakly bound electron-hole spin pairs, termed as polarons pairs (PP), prevalent in recombination currents of organic semiconductors [14]. Despite the inherently paired nature of these PPs, their weak spin-spin interaction typically allows the individual charge-carriers to behave largely as uncoupled electron spins. This unique behavior renders EDMR a preferred technique for studying drive-induced magnetic resonance phenomena [1,2,15]. In this realm, noteworthy studies are those by Ashton and Lenahan on two-photon resonances in 4H-SiC transistors [2] and by Jamali *et al.* identifying an apparent resonance at twice the field of the fundamental resonance in an organic semiconductor [15] as a two-photon transition [1]. Both studies highlighted peak shifts under strong drive but stopped short of providing a comprehensive quantitative analysis juxtaposing the observed shifts with existing theoretical descriptions.

In this paper, our focus is on understanding both the qualitative and quantitative characteristics of multi-photon magnetic dipole resonance shifts. We aim to discern whether such shifts are fundamentally similar to the well-understood shift of one-photon resonances, as detailed by Bloch and Siegert (Bloch-Siegert shift, BSS) [3], or whether they have a distinct origin. Jamali *et al.* showcased that strong-drive processes can be meticulously described via a non-perturbative approach using Floquet-analysis of the time-dependent Hamiltonian [1]. Their numerical Monte Carlo simulations exhibited qualitative congruence with experimental EDMR spectra, shedding light on several strong-drive magnetic resonance phenomena. Yet, due to calibration limits for the static magnetic field, $B_0$, and potential interference from overlapping one-photon spin resonances induced by higher harmonics of the RF amplifier, as elaborated upon previously [15], a comprehensive quantitative analysis of the subtle peak-center shifts has been lacking, leaving a gap in the validation of existing theoretical predictions. Building on the approach by Jamali *et al.*, we provide an explanation for the experimentally observed results.

## 2. Hypothesis: Non-BSS-type resonance shifts

Using the numerical simulation tool reported by Jamali *et al.* [1], we first consider predictions of changes in recombination current from the steady state, at room temperature and under constant bias, induced by weakly dipole- and exchange-coupled spin-1/2 electron-hole charge-carrier pairs in OLEDs. The active layers of these devices are made from the perdeuterated π-conjugated polymer poly[2-methoxy-5-(2'-ethylhexyloxy)-1,4-phenylene vinylene] (d-MEH-PPV). This material has demonstrated suitability for low magnetic-field (i.e., low frequency) EDMR [16] and, thus, was chosen as a model system for the detection of electron paramagnetic resonances in the strong coupling limit.

Figure 2 displays a comparison between Floquet-theory based simulations of OLED current as a function of $B_0$ and $B_1$ for (a) linearly and (b) circularly polarized strong RF excitation. The numerical data shows that the one-photon resonance shift observed under linear excitation disappears under circularly polarized (CP) excitation. In contrast, the two-photon shift remains nearly unchanged for both linearly polarized (LP) and CP excitation, with a small difference indicated by the two arrows in each plot. As predicted by Bloch and Siegert, the one-photon shift requires the presence of a counter-rotating, off-resonant component of $B_1$ [3], which is present in an LP field but not in a purely CP field. A strong shift of the two-photon resonance is seen irrespective of polarization of $B_1$ and consequently motivates the central hypothesis of this study, namely that the underlying mechanism of the two-photon resonance shift is different from that of the BSS.

Simple polarization-based arguments show that in two-level systems under magnetic resonance with the static field $\boldsymbol{B}_0$ perpendicular to the drive field $\boldsymbol{B}_1$, only odd-photon transitions may occur, while even-photon transitions become possible when the angle between $\mathbf{B}_0$ and $\mathbf{B}_1$ is tilted away from 90° [16]. A thorough analysis of the tilt-angle dependence of one- and two-photon transitions was given in Ref. [6], based on Floquet theory [4]. Below, we extend the analysis of Ref. [6] to multi-photon transitions under arbitrary drive-field helicity, emphasizing the principal difference between the one-photon (Bloch-Siegert) resonance shift and the two-photon resonance shift. We then scrutinize those expressions with numerical simulations based on the work of Jamali *et al.* [1] as well as with our recent experimental data.

### 3. Analytical expressions for multi-photon magnetic resonance shifts and intensities

We consider a single spin $S = 1/2$ in a hyperfine magnetic field $\mathbf{B}_\mathrm{hf}$ (that results from unresolved hyperfine couplings between the PP electronic spins and the surrounding nuclear spins) and externally applied static and RF magnetic fields, $\mathbf{B}_0 = B_0 \hat{\mathbf{z}}$ and $\mathbf{B}_\mathrm{RF} = B_1 \cos(\omega t)\hat{\mathbf{x}}$. To simplify the calculations, we align the quantization axis $\hat{\mathbf{z}}$ with the total static field, $\mathbf{B}_0 + \mathbf{B}_\mathrm{hf}$, and place $\mathbf{B}_\mathrm{RF}$ in the $xz$-plane. The angle $\theta$ between the new $x$-axis and the driving field $\mathbf{B}_\mathrm{RF}$ is expressed through the original $x$-component of the hyperfine field as $\sin\theta = B_{\mathrm{hf},x}/|\mathbf{B}_0 + \mathbf{B}_\mathrm{hf}|$. The energies of the two Zeeman-split levels are $E_\alpha = -\gamma|\mathbf{B}_0 + \mathbf{B}_\mathrm{hf}|/2$ and $E_\beta = \gamma|\mathbf{B}_0 + \mathbf{B}_\mathrm{hf}|/2$, where $\gamma$ is the gyromagnetic ratio. The Floquet Hamiltonian [4] in the basis of dressed Floquet states, $\ldots, |\alpha, -1\rangle, |\beta, -1\rangle, |\alpha, 0\rangle, |\beta, 0\rangle, |\alpha, 1\rangle, |\beta, 1\rangle, \ldots$, is

$$\begin{pmatrix}
\cdot & \cdot & \cdot & \cdot & \cdot & \cdot & \cdot & \cdot \\
\cdot & E_\alpha - \omega & 0 & r_{11} & r_{12} & 0 & 0 & \cdot \\
\cdot & 0 & E_\beta - \omega & r_{21} & r_{22} & 0 & 0 & \cdot \\
\cdot & r_{11}^* & r_{21}^* & E_\alpha & 0 & r_{11} & r_{12} & \cdot \\
\cdot & r_{12}^* & r_{22}^* & 0 & E_\beta & r_{21} & r_{22} & \cdot \\
\cdot & 0 & 0 & r_{11}^* & r_{21}^* & E_\alpha + \omega & 0 & \cdot \\
\cdot & 0 & 0 & r_{12}^* & r_{22}^* & 0 & E_\beta + \omega & \cdot \\
\cdot & \cdot & \cdot & \cdot & \cdot & \cdot & \cdot & \cdot
\end{pmatrix} \quad (1)$$

Introducing CP co-rotating and counter-rotating components of RF, $B_1$ and $\bar{B}_1$, the off-diagonal elements of Eq. (1) are

$$r_{11} = \frac{1}{8}\gamma(B_1 + \bar{B}_1)\sin\theta,$$

$$r_{12} = \frac{1}{4}\gamma\left[\bar{B}_1\cos^2\frac{\theta}{2} - B_1\sin^2\frac{\theta}{2}\right],$$

$$r_{21} = \frac{1}{4}\gamma\left[B_1\cos^2\frac{\theta}{2} - \bar{B}_1\sin^2\frac{\theta}{2}\right], \quad (2)$$

$$r_{22} = -r_{11}.$$

Typically, the tilt angle $\theta$ is small, so that $r_{12}$ predominantly corresponds to the counter-rotating component of RF, and $r_{21}$ to the co-rotating component. The diagonal elements $r_{11}$ and $r_{22}$ represent a component of RF that is parallel to the quantization axis due to the non-zero tilt.

We now consider the situation where the applied frequency $\omega$ is nearly resonant with the energy separation of the two states, $E_\beta \approx E_\alpha + \omega$. Following Shirley [4] we separate the portion of the Hamiltonian Eq. (1) acting on the two nearly degenerate levels as a two-by-two matrix, $\mathcal{H}_2$, and account for the rest of the Hamiltonian approximately by including perturbation corrections to the two nearly degenerate levels and coupling elements thereof from the remaining states. Specifically, we separate the nearly degenerate levels $|\beta, 0\rangle$ and $|\alpha, 1\rangle$ that are non-resonantly coupled with the states $|\alpha, -1\rangle$ and $|\beta, 2\rangle$, respectively. This coupling is given by $r_{12}$, dominated by the counter-rotating component of the RF. Incorporating this coupling into $\mathcal{H}_2$ by perturbation theory gives

$$\mathcal{H}_2 = \begin{pmatrix} E_\beta + \delta_\beta & r_{21} \\ r_{21}^* & E_\alpha + \delta_\alpha + \omega \end{pmatrix}, \qquad (3)$$

where

$$\delta_\alpha = \frac{|r_{12}|^2}{E_\alpha - E_\beta - \omega} \simeq -\frac{|r_{12}|^2}{2\omega}, \quad \delta_\beta = -\delta_\alpha. \qquad (4)$$

Note that the non-resonant interactions among the states $|\alpha, n\rangle$ through $r_{11}$, as well as those among $|\beta, n\rangle$ through $r_{22}$, cancel out to the leading order.

The resonance frequency from Eq. (4) is

$$\omega_{\text{res}} = E_\beta + \delta_\beta - E_\alpha - \delta_\alpha \simeq E_\beta - E_\alpha + |r_{12}|^2/\omega. \qquad (5)$$

The last term in Eq. (5) is the RF-induced resonance shift, i.e., the BSS,

$$\delta\omega_{\text{res}} = |r_{12}|^2/\omega. \qquad (6)$$

It is clear that the BSS is induced mostly by the counter-rotating component $\bar{B}_1$ of the RF field and disappears if $\bar{B}_1 = 0$ and $\theta = 0$.

Now we consider the energy separation of the two states close to the two-photon resonance; $E_\beta \approx E_\alpha + 2\omega$. In this case, separating the portion of the Hamiltonian Eq. (1) involving the nearly degenerate levels (i.e., finding the corresponding $\mathcal{H}_2$) is less straightforward, as the resonating pairs of states, such as $|\beta, -1\rangle$ and $|\alpha, 1\rangle$, are not directly coupled and are non-resonantly coupled to the nearest intermediate states $|\alpha, 0\rangle$ and $|\beta, 0\rangle$ in addition to the *outer* states $|\alpha, -2\rangle$ and $|\beta, 2\rangle$. We find

$$\mathcal{H}_2 = \begin{pmatrix} E_\beta - \omega - \delta_{2p} & u \\ u^* & E_\alpha + \omega + \delta_{2p} \end{pmatrix}, \tag{7}$$

where

$$\begin{aligned} \delta_{2p} &= \frac{|r_{12}|^2}{E_\alpha - E_\beta - \omega} + \frac{|r_{21}|^2}{E_\alpha - E_\beta + \omega} \\ &\simeq -\frac{|r_{12}|^2}{3\omega} - \frac{|r_{21}|^2}{\omega}, \\ u &= \frac{r_{21} r_{22}}{E_\alpha - E_\beta + \omega} + \frac{r_{11} r_{21}}{\omega} \\ &\simeq \frac{r_{21}(r_{11} - r_{22})}{\omega}. \end{aligned} \tag{8}$$

The resonance frequency of the two-photon transition is thus

$$\begin{aligned} \omega_{\text{res},2p} &= \frac{E_\beta - E_\alpha - 2\delta_{2p}}{2} \\ &\simeq \frac{E_\beta - E_\alpha}{2} + \frac{|r_{12}|^2}{3\omega} + \frac{|r_{21}|^2}{\omega}. \end{aligned} \tag{9}$$

In contrast to the BSS of the fundamental resonance, Eq. (6), the two-photon resonance shift has contributions from both co- and counter-rotating components of the RF field,

$$\delta\omega_{\text{res},2p} = \frac{|r_{12}|^2}{3\omega} + \frac{|r_{21}|^2}{\omega}. \tag{10}$$

Moreover, as seen from Eq. (10), in the case of a LP RF field, only 1/4 of the shift comes from the counter-rotating component (i.e., $r_{12}$) whereas 3/4 of the shift comes from the co-rotating component ($r_{21}$).

In fact, Eq. (8) is universal and describes the leading-order correction to the Floquet energy levels near all higher harmonic, $n$-photon resonances for $n \geq 2$. In analogy with Eq. (9), the $n$-photon resonance occurs when the $n$-tuple of the frequency is around the energy separation, $n\omega \approx E_\beta - E_\alpha$, or more precisely, at

$$\omega_{\text{res},np} = \frac{E_\beta - E_\alpha - 2\delta_{np}}{n}. \tag{11}$$

Utilizing $E_\beta - E_\alpha = n\omega$ in Eq. (8), the relation for the shift of the $n$-photon resonance can be expressed as

$$\delta\omega_{\text{res},np} = \frac{2}{n}\left[\frac{|r_{12}|^2}{(n+1)\omega} + \frac{|r_{21}|^2}{(n-1)\omega}\right]. \tag{12}$$

The relations derived above can be readily reformulated into forms describing resonance positions as a function of the applied field $B_0$ for a fixed RF frequency $\omega$. Neglecting the contribution of the hyperfine field to energy separation and writing $E_\beta - E_\alpha = \gamma B_0$, the one-photon resonance takes place at

$$\begin{aligned}\gamma B_0 &= \omega + 2\frac{|r_{12}|^2}{E_\alpha - E_\beta - \omega} \\ &\simeq \omega - \frac{|r_{12}|^2}{\omega}.\end{aligned} \quad (13)$$

For the $n$-photon transition ($n \geq 2$), the resonance condition is

$$\begin{aligned}\gamma B_0 &= n\omega + 2\left[\frac{|r_{12}|^2}{E_\alpha - E_\beta - \omega} + \frac{|r_{21}|^2}{E_\alpha - E_\beta + \omega}\right] \\ &\simeq n\omega - 2\left[\frac{|r_{12}|^2}{(n+1)\omega} + \frac{|r_{21}|^2}{(n-1)\omega}\right].\end{aligned} \quad (14)$$

These expressions elucidate the helicity dependence of the resonance line shifts. In any resonant transition, the upper energy spin state couples non-resonantly to the Floquet state above it through the counter-rotating component of $B_1$, while the lower energy state couples to the Floquet state below it through the same counter-rotating component. In single-photon transitions, the co-rotating component directly couples the two spin states. In multi-photon transitions, however, the spin states are separated by one or more intermediate Floquet states, and the co-rotating component couples each spin state non-resonantly to the nearest intermediate Floquet state. For an LP field, i.e., $\bar{B}_1 = B_1$, the intermediate Floquet states are energetically closer to the spin states than the outer Floquet states and thus exert a stronger influence, so that multi-photon resonance lines depend more strongly on the co-rotating than on the counter-rotating component. In contrast, the shift of the one-photon resonance, which has no intermediate Floquet states, depends only on the counter-rotating component. These results are consistent with the numerical data in Figure 2. Furthermore, Eq. (14) predicts that as $n \to \infty$ in an $n$-photon transition, i.e., as the energy difference between the upper and lower spin states becomes very large compared to the difference from their nearest outer and intermediate Floquet states, the contributions to the shift from both the co- and counter-rotating components approach equal values and also approach zero.

Using Eq. (2) with $\bar{B}_1 = B_1$ (LP RF field), we get

$$\begin{aligned}B_0 &\simeq \frac{\omega}{\gamma} - \frac{\gamma B_1^2 \cos^2\theta}{16\omega}, \quad \text{for } n = 1, \\ B_0 &\simeq n\frac{\omega}{\gamma} - \frac{\gamma B_1^2 \cos^2\theta}{4\omega}\frac{n}{n^2-1}, \quad \text{for } n \geq 2,\end{aligned} \quad (15)$$

implying that for any LP amplitude of $B_1$ in the strong-drive regime, the ratio of the shift of an $n$-photon resonance ($n \geq 2$) and the one-photon resonance must be

$$\frac{\Delta B_{\gamma_n}}{\Delta B_{\gamma_1}} = \frac{n\frac{\omega}{\gamma} - B_0}{\frac{\omega}{\gamma} - B_0}$$

$$= \frac{\frac{\gamma B_1^2 \cos^2\theta}{4\omega} \frac{n}{n^2 - 1}}{\frac{\gamma B_1^2 \cos^2\theta}{16\omega}} \quad (16)$$

$$= \frac{4n}{n^2 - 1}$$

This result depends only on the number of photons in the $n$-photon resonance being tested. For the two-photon shift, the predicted ratio is $8/3 \approx 2.67$.

## 4. Analysis of numerical data

In order to verify the consistency of the result given by Eq. (16) with the numerical predictions by the Floquet theory described above, we compared the result to resonance line-shift ratios obtained from the $B_0$ values of peak extrema as a function of $B_1$. The results of this procedure are displayed in Figure 5(b), together with experimental data that are discussed below. Due to the absence of analytical expressions for the individual EDMR resonances that emerge in the strong-drive regime, we analyze the simulated data on the basis of local extrema of the EDMR spectra, with the understanding that local extrema caused by resonance lines do not always coincide with resonance-line centers, e.g. when line shapes are asymmetric or resonance lines overlap. Vertical uncertainty intervals originate from inaccuracies in the magnetic field where local extrema occur, analogous to error bars of experimental data, and were obtained by fitting the simulated spectra with second-order polynomials, taking the standard deviation of the residuals, and then fitting the peaks again using the standard deviation to generate a covariance matrix, from which the error in the resonance-line center was computed. For d-MEH-PPV, this method leads to a poor definition of the one-photon peak centers between $B_1 \sim 0.3$ mT and $0.5$ mT because the resonance peak inverts due to the onset of spin-collectivity—the so-called spin-Dicke effect [17]—and a local extremum becomes undefined, producing large uncertainty intervals for the resonance-line center and the resonance-line shift at $B_1 \sim 0.35$ mT. This transition causes a non-monotonic interval in the resonance line shift versus $B_1$ [cf. Figure 5(a)].

In preparation for the experimental work, we also studied numerically whether the extrema-based analysis of resonance line-shift data could be affected by the modulation envelope used for lock-in detection. The theoretical treatment of lock-in measurements with sinusoidal modulation of the RF amplitude, which was the experimental configuration, is computationally expensive compared to the treatment of rectangular modulation. In the case of rectangular modulation, the EDMR signal under slow modulation (typically $f_s \lesssim$ 1 kHz in our experiments) is proportional to the difference of steady-state OLED currents with and without the RF drive of amplitude $B_1$, i.e.,

$$\text{EDMR}(B_0) \propto I(B_1, B_0) - I(0, B_0). \tag{17}$$

In this case, the EDMR spectra may be calculated using the truncation scheme described in Ref. [4] by restricting the Floquet degree of freedom to a finite domain, $-N_0 \leq n \leq N_0$. Thus, the truncated Floquet Hilbert space is $4(2N_0 + 1)$-dimensional. The optimal value of $N_0$ is found by inspecting the convergence of the simulation result against increasing $N_0$. We have verified numerically that $N_0 = 4$, corresponding to a 36 × 36 truncated Floquet Hamiltonian, ensures an acceptable convergence for $I(B_1, B_0)$ in the parameter domain of interest. This approach provides a convenient numerical procedure for the calculation of $I(B_1, B_0)$ on a standard desktop processor within a reasonable time frame. In contrast to rectangular modulation, sinusoidal modulation does not allow for a simple interpretation of the EDMR spectra as in Eq. (18). The RF amplifier output voltage is described by

$$V = V_0 \cos(2\pi f_f t)(1 + m\cos(2\pi f_s t)), \tag{18}$$

where the carrier frequency $f_f = 100$ MHz and the modulation frequency $f_s = 1$ kHz, while $m = 0.9$ is the modulation depth. To simulate the corresponding spectra, we decompose the signal described by Eq. (19) into

$$B_1(t) = B_1\left[\cos(2\pi f_f t) + \frac{m}{2}\cos(2\pi(f_f + f_s)t) + \frac{m}{2}\cos(2\pi(f_f - f_s)t)\right], \tag{19}$$

and consider a system subject to a multi-frequency drive involving three asynchronous incident RF fields of frequencies $f_f$, $(f_f + f_s)$, and $(f_f - f_s)$. The extension of the Floquet theory for this multi-frequency drive, Eq. (20), is straightforward. An additional Floquet-degree of freedom is introduced, and the dressed states are expressed as $|\alpha, n_1, n_2\rangle$, with two integers $n_1$ and $n_2$ that represent the $f_f$- and $f_s$-photon numbers. The truncation is realized by restricting the range of the two integers, $-N_f \leq n_1 \leq N_f$ and $-N_s \leq n_2 \leq N_s$, and results in a $4(2N_f + 1)(2N_s + 1)$-dimensional Hilbert space. As above, the optimal values of $N_f$ and $N_s$ are found by testing the convergence of the simulated EDMR spectra. Our simulations show that the necessary convergence is achieved if $N_0, N_m \geq 4$. Hence, the truncated Hilbert space is at least $4 \cdot 9^2 = 324$-dimensional. Because of this large size, the numerical simulation of EDMR lines for sinusoidal RF modulation requires extraordinary computational cost, compared to simulation for rectangular modulation ($\geq 36$-dimensional). Thus, although our simulations reveal that these two modulation modalities impact

EDMR line shapes, we found that the difference between peak centers (mean 6 µT, standard deviation 20 µT, $n = 3$) is less than the uncertainty limits of the experimental data (mean 17 µT, standard deviation 15 µT, $n = 30$) and, thus, we focus the discussion of numerically simulated EDMR lines on rectangular RF modulation, using Eq. (18) rather than Eq. (20). Furthermore, the conclusions drawn from experimental results for resonance-line shift measurements apply equally to both experiments conducted with sinusoidal and rectangular modulation.

Concluding this discussion, we find agreement between Eq. (16) and the numerical simulations, from which resonance-line shifts were determined through the identification of local extrema and this agreement validates the procedure described above as a viable analysis technique for the experimental data that is discussed next.

## 5. Experimental test of Floquet theory predictions

The description of the strong-drive regime with Floquet theory, as given in Ref. [1], offers quantitative numerical predictions for single- and multi-photon line shifts and their dependencies on the drive field amplitude $B_1$. While precise control of $B_1$ during measurements should substantiate the accuracy of these predictions, there are technical challenges. Even with EDMR-detected, spin-dependent currents, which allow access to electron spin-resonance conditions where $B_0 \sim B_1$ [1,2,15,17], resonance shifts induced by drive power are small. They are not only small in comparison to the Zeeman splitting of the charge carrier spin, but also relative to the resonance-line widths dominated by random hyperfine fields of normally protonated organic semiconductors [18]. To mitigate the latter, we used isotopic substitution, that is, we employed a perdeuterated conjugated polymer for the key results of the study.

Another challenge arises from the continuous wave (c.w.) high-intensity RF excitation, which leads to radiation-induced artifact signals adding to the uncertainty in determining the resonance line shifts. Still another difficulty is determining the experimental value of $B_1$. Given that sample alignment, the resonator quality factor, and the coupling between sample and radiation field are sensitive to small positional changes of the sample relative to the excitation coils and sample holder, the conversion factor between the square root of the applied RF power and $B_1$ can change dramatically between experimental runs, when the sample holder is removed from the magnet and a new sample placed into it. The determination of $B_1$ is therefore most reliably done during the recording of a c.w. EDMR spectrum. This, however, conflicts with the need to conduct transient spin nutation measurements (the standard way of measuring $B_1$), which are pulsed EDMR-detected spin-Rabi oscillation experiments [17] and therefore cannot be conducted simultaneously with c.w. EDMR experiments. To address this problem, we adopted two strategies: (i) we considered the resonance-line shifts as functions of $B_1$ alone and then calibrated the RF power scale to $B_1$ through an analysis of power broadening in the intermediate-drive regime, following the procedure described by Jamali *et al*. [15]; and (ii) we compared the two-photon resonance line shifts to the BSS of the one-photon resonance. Given that the BSS of the one-photon electron spin resonance has a well-understood dependence on $B_1$, the numerical Floquet theory predictions about the dependence of the two-photon shift on $B_1$ can be evaluated by comparing their agreement after the one-photon shifts are scaled along the $B_1$ axis to match *their* respective Floquet predictions [cf. Figure 5(a)]. This method avoids the need to have precise $B_1$ values. In contrast, Eq. (16) allows us to test the Floquet analysis experimentally without any knowledge of $B_1$, yet it does require EDMR spectra to be measured with an absolute magnetic-field ($B_0$) scale that allows accurate quantification of the power-induced one- and two-photon resonance shifts. In addition, the EDMR current detection chain must allow for strong suppression of spurious signals, e,g., those caused by one-photon resonances of higher amplifier harmonics [15], and for improved overall signal-to-noise ratio (SNR) compared to our previously reported strong-drive EDMR experiments on d-MEH-PPV [1].

## 6. Experimental methods and results

Figure 3(a) depicts the experimental setup. The circuit that generates $B_0$ consists of a Varian V3603 electromagnet, a Kepco 20-20D bipolar power supply, and a National Instruments PI 6221 multifunction I/O device. The magnetic flux between the poles of the electromagnet is measured using a Hall sensor connected to an F.W. Bell Model 5080 Gauss meter. The circuit used to detect the OLED current consists of two 6 V batteries, a Stanford Research Systems SR570 low-noise current amplifier, and a Zurich HF2LI digital lock-in amplifier. The RF circuit consists of a Hewlett Packard 8656B signal generator (modulated by the lock-in amplifier), an ENI 5100L RF amplifier, a Microwave Filter Company 17842-6 low-pass filter (passband 0-120 MHz, nominal 0.5 dB insertion loss, and minimum 50 dB rejection between 127 and 488.5 MHz), a custom-made copper coil surrounding the OLED, a Fairview ST3NF50PL 50 Ω termination resistor, and a series sensing resistor consisting of two parallel 1 Ω resistors. The waveform across the sensing resistor was recorded through an Agilent MSO6104A oscilloscope.

The d-MEH-PPV-based OLEDs were fabricated as previously reported [1,14,19,20] [cf. Figure 4(c)]. Data were collected from two different samples, labeled Sample 1 and Sample 2. Sample 1 had a larger active area (5 mm$^2$) than Sample 2 (0.8 mm$^2$) and thus a better SNR, while the smaller area of Sample 2 improved heat sinking and homogeneity of the applied magnetic fields across the device's active area [21], all of which were advantageous in experiments at high RF-drive fields.

We implemented an RF-amplitude modulation scheme of $B_1$ with phase-sensitive (i.e., lock-in) detection of the sample current, allowing for the separation of the magnetic-resonance-induced, spin-dependent current from radiation-induced, spin-independent electrical currents. The latter currents contribute random artifact signals, yet they are not strictly noise because they follow the RF-amplitude modulation and therefore cannot be filtered out solely by the narrow bandpass filter behavior of a lock-in amplifier. Nevertheless, it is possible to separate these electrical signals from spin-dependent electric current signals using a lock-in detector because of their entirely different dynamical signatures. The dynamics of spin-dependent recombination occur at kilohertz to megahertz rates, while the radiation-induced electronic effects are much faster, in the megahertz to gigahertz range. Thus, the two signal types can be separated between the in-phase ($X$) and quadrature ($Y$) channels by suitable choice of modulation frequency and reference phase. The radiation-induced signal was typically minimized in the $Y$ channel to below the noise limit of the spin-dependent recombination currents. We recorded all spectra using a fixed reference phase and then digitally adjusted it to minimize the spin-independent contributions in the out-of-phase ($Y$) channel, as depicted in Figure 3(b).

After adjusting the reference phase, the experimental spectra were analyzed according to the following procedure: first, the signal was subjected to a linear baseline correction to compensate for non-resonant, RF-induced, quasi-static magnetoresistance effects [22]. Second, the EDMR peak centers were determined by second-order polynomial fits around the local extrema. The error reported for each extremum propagated through the covariance matrix of the fit and originated from the noise distribution of the ordinates of the spectra. Third, the peak centers of the one-photon resonance at low resonant drive fields were used to calibrate $B_0$, following the previously reported robust absolute magnetometry approach based on spin-dependent recombination [23]. During the experiment, we swept the magnetic field $B_0$ bi-directionally (i.e., recording the EDMR trace for two opposite magnetic-field directions) in order to obtain symmetric one-photon peak centers. We then extracted a linear function consisting of a scaling factor $a$ and an offset constant $b$, with $a$ being calculated from the one-photon resonance peaks in the low RF-power regime, where the peak centers align with the theoretical Zeeman energy of a free electron, $\omega/\gamma$, and the offset $b$ being determined from the midpoints of the one-photon EDMR peaks. This approach allowed us to obtain a highly accurate calibration of the magnetic field, i.e., the relationship between the value of $B_0$ measured by the Hall sensor and the actual values of $B_0$.

We recorded 1) room temperature changes to a ~10 µA steady-state forward current of a d-MEH-PPV-based OLED, 2) the applied magnetic field $B_0$, and 3) the voltage across the sensing resistor connected in series with the RF coil, as a measure of $B_1$. This detection scheme follows those of previous experiments [1,15,17], but with the following three changes: (i) the RF-amplitude modulation and lock-in detection scheme described above was implemented; (ii) we conducted bi-directional field sweeps, also described above; and (iii) a low-pass filter was inserted between the output of the RF amplifier and the RF coil, similar to approaches previously reported [7,8]. As the high power required to produce large c.w. $B_1$ pushes RF amplifiers into the non-linear response domain, higher RF harmonics can arise, requiring RF filtering. To test the ability of our c.w. EDMR setup to suppress resonances caused by higher amplifier harmonics, we recorded EDMR spectra with OLEDs fabricated with commercially available super-yellow (SY) PPV [cf. Figure 4(c)], as described in Ref. [15]. The measured data, shown in Figure 4(a,b), were obtained under strong drive amplitudes ($B_1$~0.5 mT) (a) without and (b) with the inclusion of an RF low-pass filter ($f_c$ ~ 120 MHz). SY PPV has stronger local hyperfine fields than d-MEH-PPV due to it being fully protonated. Consequently, for any given value of $B_1$, a much weaker two-photon shift arises than in d-MEH-PPV [17,24]. In Figure 4(b), the features at the second harmonic around $|B_0|$~7 mT are substantially reduced compared to panel (a), thereby demonstrating the possibility to detect the two-photon resonance without the superimposed contribution of higher RF harmonics. In contrast, the features seen at $|B_0|$~10 mT in panel (a) are completely eliminated in panel (b), thereby confirming that these features are entirely due to amplifier harmonics rather than three-photon resonances. While theory does predict the existence of three-photon resonances [1], the $B_1$ amplitude reached in these measurements was too low to observe them. We confirmed through an analysis of the power broadening of the spectra [15] that the attenuation of higher-order resonances was not due to an attenuation of $B_1$.

Figure 4(d) shows several spectra recorded at high RF power with the d-MEH-PPV OLEDs, with the baseline correction applied as well as the $B_0$ calibration. The spectra are layered from top to bottom in order of increasing $B_1$, which was estimated by assuming a linear relationship $B_1 = cV$ between $B_1$ and the RF amplitude $V$. The conversion factor $c$ was determined through a global fit to several spectra in the intermediate-drive regime as described in Ref. [15]. We accounted for the effect of sinusoidal modulation (Eq. (19)) on linewidth by assuming that the system reached a steady state on a timescale much shorter than the modulation period $1/f_s$. Under that assumption, the linewidth modulation averages to zero over the modulation period, so the time-averaged modulated linewidth can be treated as the unmodulated linewidth. The values of $c$ were calculated separately for both Sample 1 and Sample 2 and were then averaged because both values were within two standard deviations, which were calculated for Sample 1 from the statistics of several fits and for Sample 2 through a numerical bootstrapping method [18,25]. In Figure 4(d), the monotonic increase of both the inversion of the one-photon resonance peak—a signature of the spin-Dicke effect [17]—and of the two-photon resonance intensity with increasing $B_1$ agrees qualitatively with the data reported previously [1] and thus validates the analysis procedure described above to estimate $B_1$. Nevertheless, the value of $c$, which was determined from spectra recorded in the intermediate-power RF regime, is expected to have changed at high power as Ohmic heating effects shifted the overall impedance and tuning of the RF circuit. This uncertainty restricts the resulting values of $B_1$ from being used as an absolute scale for a quantitative test of Floquet theory. Finally, we note that the spectra in Figure 4(d) are only a subset of all the recorded data, while additional datasets were used for the analysis of the resonance-line shifts discussed below.

## 7. Data analysis and discussion

Figure 5(a) shows the values of $B_0$ where local extrema occurred, corresponding to the one- and two-photon resonance centers, for both the experimental and numerical data, plotted as a function of $B_1$. The horizontal ($B_1$) scale was determined by the optimal alignment of the experimental one-photon resonance peaks with the numerical one-photon resonance peaks, the correction factor being determined through a numerical least-squares fitting procedure. As seen in Figure 5(a), the corrected $B_1$ scale shows excellent agreement not only between the one-photon experimental and numerical values, but also between the two-photon values. As outlined in Section 5, the agreement between the experimental and numerical two-photon peak shifts, after aligning with the well-understood one-photon peak shifts, confirms the underlying Floquet model. While the experimental results from Sample 1 and Sample 2 cover different ranges of $B_1$, all the data consistently exhibit resonance line center shifts with increasing $B_1$.

Given the uncertainty of the absolute scale of $B_1$ discussed above, we also examined the ratio of one- and two-photon resonance line shifts ratio predicted by Eq. (16), which is independent of $B_1$. Figure 5(b) plots the two-photon shift as a function of the one-photon shift for each experimental and numerical spectrum. Remarkably, notwithstanding that the shifts of the one- and two-photon peak centers have a highly non-linear dependence on $B_1$, they are mutually linear, with a ratio of $8/3 \approx 2.67$, in excellent agreement with the prediction of Eq. (16).

The experimental data and quantitative analysis presented here agree with the analytical and numerical predictions from Floquet theory under linear polarization and, by extension, support the hypothesis that the underlying mechanism of the two-photon resonance shift is different from that of the BSS. Beyond these results, the analytical expressions call for an experiment with arbitrarily controllable drive-field helicities, as well as an experiment with arbitrary control of the angle $\theta$ between $B_1$ and $B_0$. Furthermore, for additional experimental scrutiny, it is desirable in EDMR and ODMR experiments to find methods to produce higher $B_1/B_0$ ratios.

## 8. Summary and Conclusions

In summary, we derive analytical expressions showing that multi-photon resonance line shifts depend on both the co- and counter-rotating components of the drive-field, allowing for experimental scrutiny of the Floquet theory of strong-drive magnetic dipole transitions. We report the experimental observation of power-induced shifts in one- and two-photon magnetic-dipole resonances in EDMR at low magnetic field and low frequency. The experiments, which utilized spin-dependent recombination currents through polaron-pair states in OLEDs with perdeuterated MEH-PPV as the active layer, confirm the agreement between analytical expressions and numerical simulations under linear drive-field polarization and suggest that the observed two-photon magnetic-dipole transitions are indeed only weakly helicity dependent. Our results motivate further studies of the two-photon-induced resonance shift behavior under strong magnetic-resonant drive conditions—with regards to both its fundamental nature and also its potential applicability for coherent control sequences and quantum-enhanced technological applications such as spin-based quantum sensing.

## ACKNOWLEDGMENTS

This work was supported by the U.S. Department of Energy, Office of Basic Energy Sciences, Division of Materials Sciences and Engineering under Award #DE-SC0000909. H. M. and V. V. M. acknowledge support from the Deutsche Forschungsgemeinschaft (DFG, German Research Foundation) (project ID 314695032 – SFB 1277, subproject B03). The synthesis of the perdeuterated monomer took place at the Australian National Deuteration Facility, which is partly funded by NCRIS, an Australian Government initiative. P.L.B. was an Australian Research Council Laureate Fellow, and the synthesis work was supported in part by this Fellowship (FL160100067).

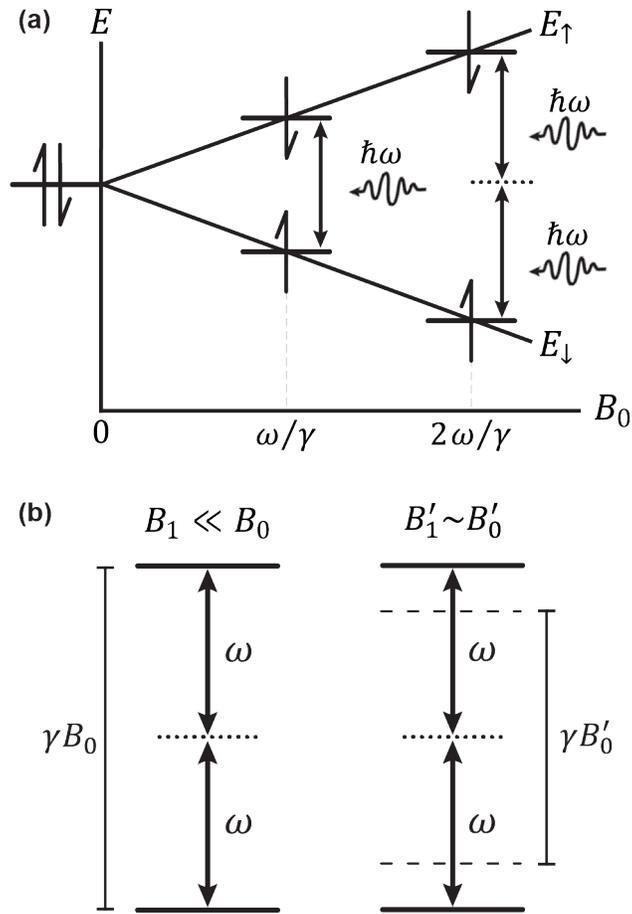

Figure 1. Conceptual representation of (a) one- and two-photon resonances, and (b) Energy level diagram in frequency units showing the RF power-induced shift of the two-photon resonance, which, under the conditions of field-swept, c.w. magnetic resonance spectroscopy, requires a reduced Zeeman splitting to maintain the resonance condition with a drive field of constant frequency. *Left*: For spin states that are weakly coupled to their environment, the eigen-energy level splitting is governed by the Zeeman effect and increases linearly with $B_0$. Thus, a two-photon magnetic resonance can occur at twice the fundamental resonance $B_0 = \omega/\gamma$. *Right*: Under strong drive conditions, where $B_1' \sim B_0'$, the energy levels are pushed further apart. The driving field frequency $\omega$ thus matches the level spacing at a reduced field $B_0' < B_0$, so that the two-photon resonance line shifts toward $B_0 = 0$.

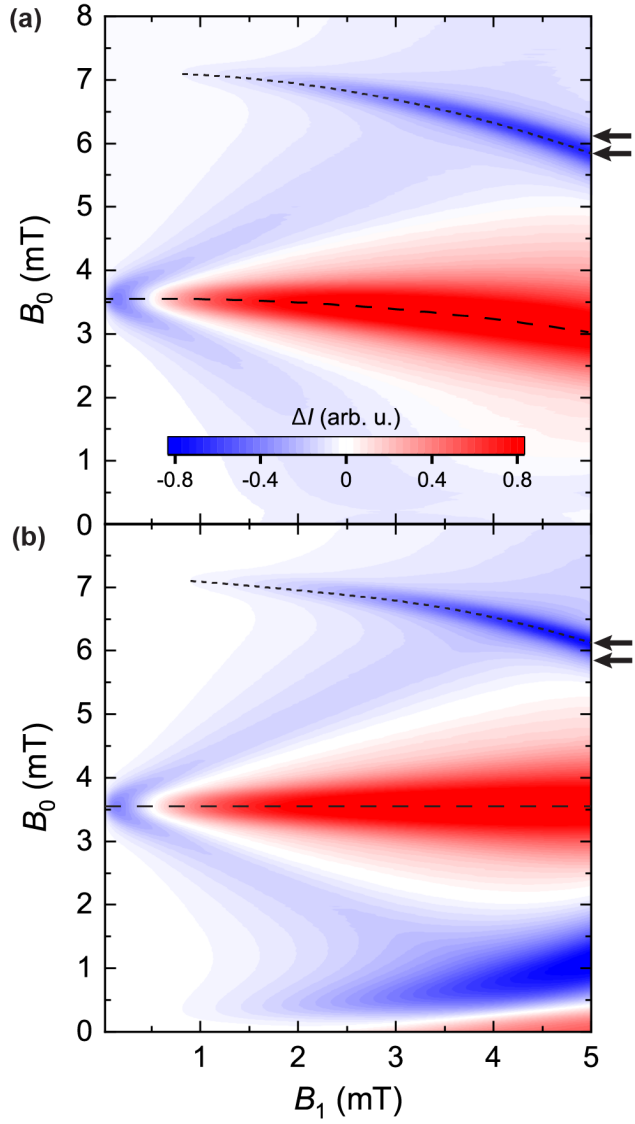

Figure 2. Simulated change $\Delta I$ to the steady-state d-MEH-PPV OLED current as a function of $B_0$ and $B_1$ under (a) linearly and (b) circularly polarized RF field with f = 100 MHz. The dashed lines are a guide to the eye for the one-photon and two-photon resonance peak centers. The pairs of arrows are placed at identical magnetic field values in (a) and (b) and mark the difference of the two-photon peak center at $B_1 = 5$ mT under linear and circular polarization. Under linear polarization, drive-amplitude induced shifts are visible for both one- and two-photon resonances, while under circular polarization, there is no shift of the one-photon resonance, as expected for the BSS [cf. Figure 1(a)]. Remarkably, the shift of the two-photon resonance remains, albeit slightly weaker.

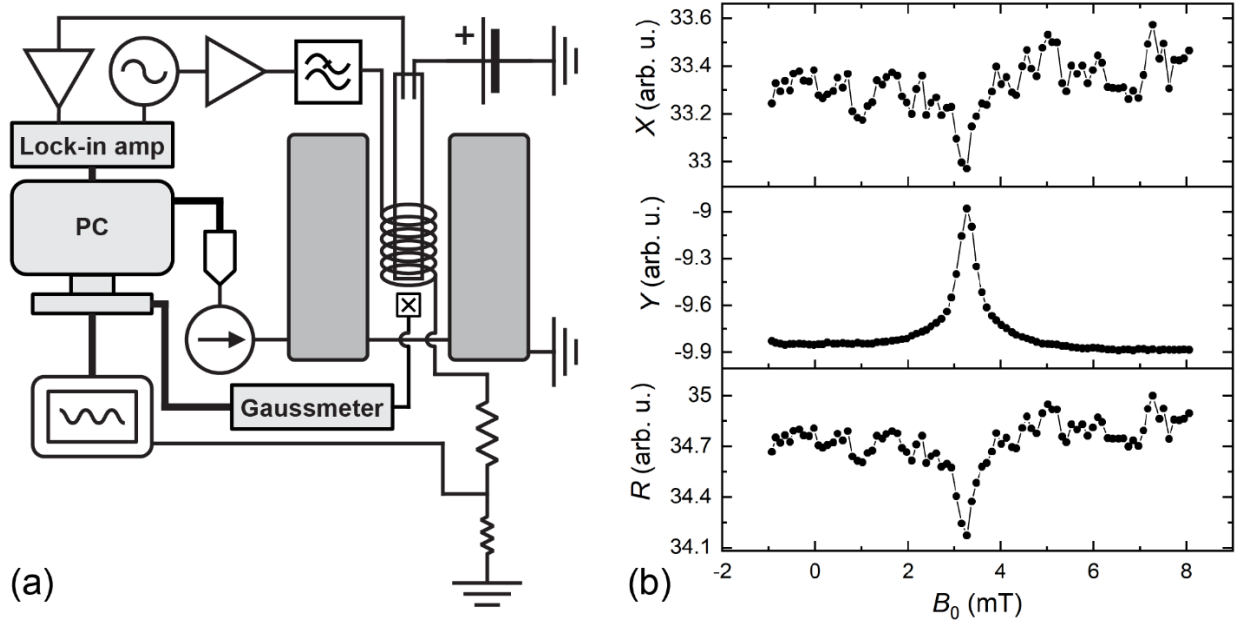

Figure 3. (a) Illustration of the experimental setup. The two thick, vertical gray bars represent the poles of the electromagnet that generates $B_0$. (b) Low-drive power EDMR spectra obtained for in-phase (top), out-of-phase (center), and the geometric sum (magnitude) of both lock-in channels. Since spurious electric signals—not related to magnetic resonance—display significantly faster dynamics, they can be separated by using a modulation frequency higher than the dominant harmonic components of the spin-dependent electric current measured here for EDMR spectroscopy, yet lower than the harmonic components of the spurious radiation artifact signals. By appropriate choice of the lock-in phase, the contributions of these spurious signals can be minimized below the noise levels within the out-of-phase channel (Y), which then solely contains EDMR signal contributions.

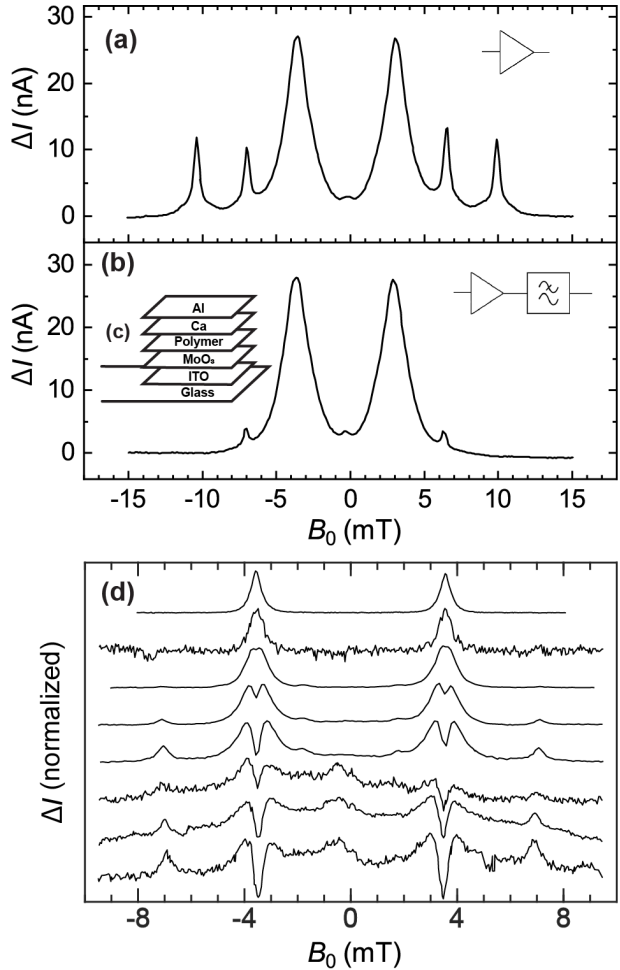

Figure 4. EDMR spectrum of a SY-PPV OLED device (a) without and (b) with an RF low-pass filter. *Inset* (c): Schematic of the OLED device. The glass substrate is (50×3×1) mm³ and has lithographically defined electrodes. (d) EDMR spectra of two different d-MEH-PPV OLED devices, based on the structure shown in (c), yet with different active areas (5 mm², 0.8 mm²), measured at high RF power increasing towards the bottom. Each spectrum is baseline-corrected, normalized, and offset on the y-axis for clarity. The scale of $B_0$ for these measurements was calibrated using the one-photon resonance at low drive fields as a standard. The measurements of the two devices took place at different, yet overlapping power intervals. Signatures of spin-collectivity (the spin-Dicke effect), indicated by the bifurcation of the one-photon resonance, as well as two-photon resonances become apparent at higher powers, i.e., at and below the fourth spectrum from the top.

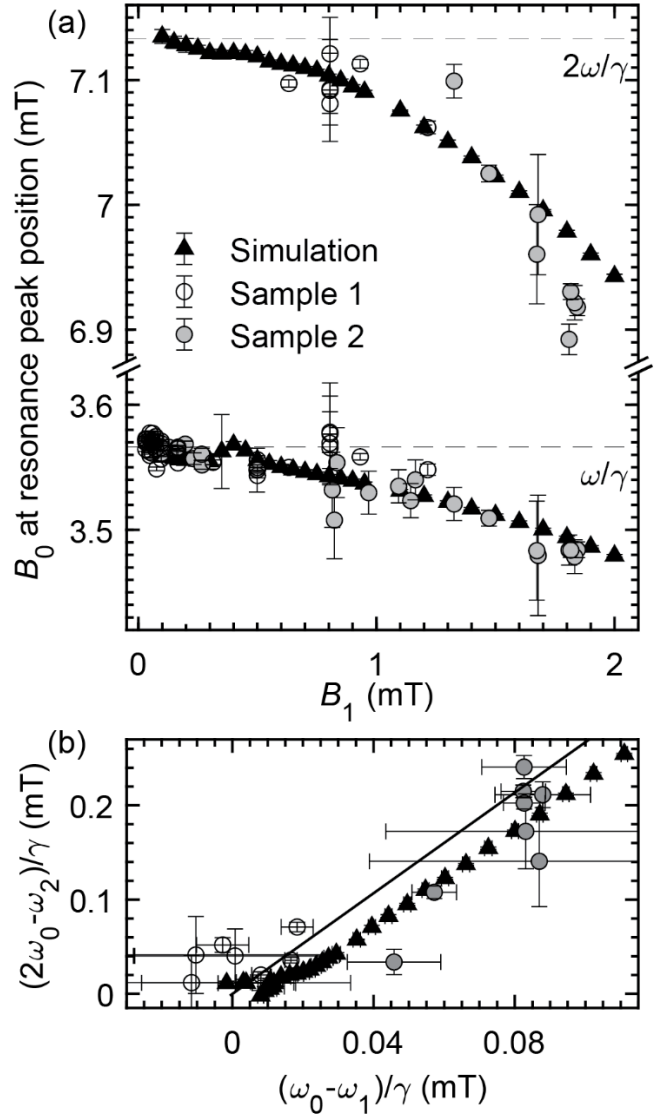

Figure 5. (a) Plots of the one- and two-photon resonance peak extrema as a function of $B_1$, on scales of $B_1$ determined by the best-fit match of the experimental and numerical one-photon peak extrema. The horizontal dashed lines mark the hypothetical unshifted peak centers. (b) Two-photon shifts as a function of one-photon shifts, using the same marker scheme as in (a). The solid line passes through the origin and has a slope of $8/3 \approx 2.67$, as predicted by Floquet theory.